\documentclass[10pt,conference]{IEEEtran}
\IEEEoverridecommandlockouts
\usepackage{cite}  
\usepackage{amsmath,amssymb,amsfonts}
\usepackage{algorithmic} 
\usepackage{textcomp}
\usepackage{xcolor}    
\usepackage{amsthm}  
\usepackage{graphicx}  
\usepackage{multirow}  
\usepackage{booktabs}   
\usepackage{subcaption} 
\usepackage{pifont}
\usepackage{enumitem}
\usepackage[linesnumbered,ruled]{algorithm2e}
\usepackage[normalem]{ulem}
\usepackage{listings}
\usepackage{balance}
\usepackage{threeparttable}

\def\BibTeX{{\rm B\kern-.05em{\sc i\kern-.025em b}\kern-.08em
    T\kern-.1667em\lower.7ex\hbox{E}\kern-.125emX}}

\makeatletter
\newcommand{\linebreakand}{%
  \end{@IEEEauthorhalign}
  \hfill\mbox{}\par
  \mbox{}\hfill\begin{@IEEEauthorhalign}
}
\makeatother

\begin{document}
\newcommand{\datalink}[1]{\textbf{https://github.com/ssmingz/VarDT}}

\newcommand{\tool}[1]{\textsc{VarDT}}
\newcommand{\variant}[1]{\tool{}$_\mathrm{\textit{#1}}$}
\newcommand{\jun}[1]{{\color{cyan}\ding{46}[Jiang: #1]}}
\newcommand{\todo}[1]{{\color{red}[#1]}}
\newcommand{\jj}[1]{{\color{orange}\ding{46}[Junjie: #1]}}
\newcommand{\ymdel}[1]{{\color{red}{\sout{#1}}}}
\newcommand{\ymadd}[1]{{\color{blue}{#1}}}

\newcommand\mycommfont[1]{\scriptsize\ttfamily\textcolor{blue}{#1}}
\SetCommentSty{mycommfont}

\newcommand{\depfactor}[1]{\textit{DEP\_FACTOR}}
\newcommand\crule[1][black]{\textcolor{#1}{\rule{5pt}{5pt}}}
\newcommand{\lin}[1]{{\scriptsize \textcolor{darkgray}{#1}}}
\lstdefinestyle{java}{ 
	language=java,
	basicstyle=\scriptsize\ttfamily, 
	breakatwhitespace=false, 
	breaklines=true, 
	captionpos=b, 
	commentstyle=\color[rgb]{0.0, 0.5, 0.69},
	deletekeywords={}, 
	escapeinside={<@}{@>},
	firstnumber=1, 
	frame=lines, 
	frameround=tttt, 
	keywordstyle={[1]\color{blue!90!black}},
	keywordstyle={[3]\color{red!80!orange}},
	morekeywords={String,int}, 
	numbers=none, 
	numbersep=-8pt, 
	numberstyle=\tiny\color[rgb]{0.1,0.1,0.1}, 
	rulecolor=\color{black}, 
	showstringspaces=false, 
	showtabs=false, 
	stepnumber=1, 
	stringstyle=\color[rgb]{0.58,0,0.82},
	tabsize=2, 
	backgroundcolor=\color{white}
}

\newcommand{\codeIn}[1]{{\ttfamily #1}}
\newcommand{\note}[1]{{\scriptsize \textcolor{darkgray}{#1}}}

\newcommand{\distance}{5pt}
\setlength{\textfloatsep}{\distance}
\setlength{\floatsep}{\distance}
\setlength{\intextsep}{\distance}
\setlength{\dbltextfloatsep}{\distance} 
\setlength{\dblfloatsep}{\distance} 

\title{Variable-Based Fault Localization via Enhanced Decision Tree}

 \author{
\IEEEauthorblockN{1\textsuperscript{st} Jiajun Jiang}
 \IEEEauthorblockA{\textit{College of Intelligence and Computing} \\
 \textit{Tianjin University}\\
 Tianjin, China \\
 jiangjiajun@tju.edu.cn}
 \and
 \IEEEauthorblockN{2\textsuperscript{nd} Yumeng Wang}
 \IEEEauthorblockA{\textit{College of Intelligence and Computing} \\
 \textit{Tianjin University}\\
 Tianjin, China \\
 jazz244008@tju.edu.cn}
 \and
 \IEEEauthorblockN{3\textsuperscript{rd} Junjie Chen\IEEEauthorrefmark{1}\thanks{\IEEEauthorrefmark{1}Corresponding author.}}
 \IEEEauthorblockA{\textit{College of Intelligence and Computing} \\
 \textit{Tianjin University}\\
 Tianjin, China \\
 junjiechen@tju.edu.cn}
 \linebreakand
 \IEEEauthorblockN{4\textsuperscript{th} Delin Lv}
 \IEEEauthorblockA{\textit{College of Intelligence and Computing} \\
 \textit{Tianjin University}\\
 Tianjin, China \\
 ldlmntq@tju.edu.cn}
 \and
 \IEEEauthorblockN{5\textsuperscript{th} Mengjiao Liu}
 \IEEEauthorblockA{\textit{College of Intelligence and Computing} \\
 \textit{Tianjin University}\\
 Tianjin, China \\
 mengjiaoliu@tju.edu.cn}
 }

\maketitle

\begin{abstract}
Fault localization, aiming at localizing the root cause of the bug under repair, has been a longstanding research topic. Although many approaches have been proposed in the last decades, 
most of the existing studies work at coarse-grained statement or method levels with very limited insights about how to repair the bug (\textit{granularity problem}), but few studies target the finer-grained fault localization. 
In this paper, we target the \textit{granularity problem} and propose a novel finer-grained variable-level fault localization technique. Specifically, we design a program-dependency-enhanced decision tree model to boost the identification of fault-relevant variables via discriminating failed and passed test cases based on the variable values. To evaluate the effectiveness of our approach, we have implemented it in a tool called \tool{} and conducted 
an extensive study over the Defects4J benchmark. The results show that \tool{} outperforms the state-of-the-art fault localization approaches with at least 247.8\% improvements in terms of bugs located at Top-1, and the average improvements are 330.5\%. 

Besides, to investigate whether our finer-grained fault localization result can further improve the effectiveness of downstream APR techniques, we have adapted \tool{} to the application of patch filtering, where \tool{} outperforms the state-of-the-art PATCH-SIM by filtering 26.0\% more incorrect patches. The results demonstrate the effectiveness of our approach and it also provides a new way of thinking for improving automatic program repair techniques.

\end{abstract}


\section{Introduction}
\label{sec:intro}


Program bugs are inevitably introduced in programs, which will potentially cause great financial losses and even disasters. Therefore, fixing bugs timely when they occur is critical. In particular, the first stage of program debugging is to locate the root cause of bugs under repair, which is an expensive and labor-intensive process. To facilitate this process, many automatic fault localization techniques have been proposed~\cite{Jones:2002:VTI:581339.581397,Tarantula,abreu2007accuracy,MUSE,Liblit-isolation,Liblit-adaptive,compound-predicate,HOLMES,Zheng:SD:multi-bug,sober,sober-TSE,Jiang-sd-cfp} in the last decades,
aiming at providing a list of candidate locations that are most possibly faulty to aid the subsequent program repair process.
%

Although great success has been achieved, the mainstream fault localization techniques still suffer from two major limitations. First, the fault localization precision is low, the state-of-the-art techniques can only locate about 21\% genuine buggy statements as the top-1 returned results~\cite{zeng2022fault}. Inaccurate fault localization results can be misleading and increase the risk of generating incorrect patches due to the incomplete specification~\cite{qi15,xiong-icse18,smith2015cure}.
Second, the granularity of existing fault localization results is still coarse-grained at statement or method levels, which provides few insights beyond locations related to the root cause for repairing the bug. As a result, even given the genuine faulty locations, the patch space is still large, which aggravates the problem of generating incorrect patches. As reported in existing studies~\cite{liu2019tbar,zhu2021syntax}, when providing genuine buggy statements, the state-of-the-art automatic program repair (APR) techniques can still repair a small number of bugs with generating many
incorrect patches, significantly affecting the usability of APR techniques. In this paper, we call these two limitations \textit{precision} and \textit{granularity} problems, respectively, in fault localization.


Over the years, the vast majority of existing studies mainly focus on the \textit{precision} problem, and have adopted different techniques, such as mutation testing~\cite{Metallaxis}, machine learning~\cite{Xuan:2014:LCM:2705615.2706097}, deep learning~\cite{TraPT,li2021fault,lou2021boosting}, etc., and incorporated diverse information like test coverage~\cite{Michael:SBFL}, program dependency~\cite{dynamic-slicing,slicing-pruning}, code changes~\cite{FLUCCS} and program invariants~\cite{B.Le2016}, to improve the precision. 
However, most of the studies work at statement or method levels, but few works target the \textit{granularity} problem, especially in the scenario of APR. Although some techniques have been designed at a finer-grained level (e.g., variable level), they are either requiring the intervention of developers~\cite{zeller-cause-effect,zeller-simplifying} or targeting a particular type of variables~\cite{kucuk2021improving,Liblit-sampling,Liblit-isolation}, making them infeasible to further the effectiveness of downstream APR techniques.



Aiming at significantly improving the effectiveness of fault localization and thus boosting the subsequent program repair process, in this paper we propose a novel and general fault localization technique, named \tool{}, targeting the \textit{granularity} problem by effectively identifying the fine-grained fault-relevant variables via leveraging a program-dependency-enhanced decision tree model. Intuitively, the basic idea of \tool{} is that fault-relevant variables may exhibit different values in failed and passed test runs, and variables that have higher discrimination ability have a larger possibility to be the root causes of the failure. 
According to this intuition, we adopt the decision tree model to aid the identification of the most fault-relevant variables by building discrimination models for failed and passed runs using candidate variables.
However, since the number of variables and their value space are usually large in real-world programs, especially in industry-grade programs, \tool{} further incorporates the static program analysis to improve its scalability and effectiveness, including program slicing and dependency analysis. We will introduce our approach detailedly in Section~\ref{sec:approach}. 

To evaluate the effectiveness of our approach, we have implemented a prototype of it as an automatic fault localization tool, also named \tool{}, and conducted an extensive experiment on the widely-used Defects4J~\cite{just2014defects4j} benchmark. The results show that \tool{} successfully located the fault-relevant variables at Top-1 position for 24.0\% of bugs, which significantly outperformed seven state-of-the-art baseline approaches. Particularly, the improvements are at least 247.8\%, and in average 330.5\% regarding the bugs located at Top-1. Moreover, to investigate whether our approach can further the effectiveness of downstream APR techniques, we also adapted \tool{} to the application of patch filtering, where it correctly filtered out 69.4\% incorrect patches. Although not designed as a comprehensive and standalone patch filtering technique, it improves the state-of-the-art PATCH-SIM by 26.0\%.
The results indicate that our finer-grained fault localization technique is indeed effective and promising to further improve the effectiveness of downstream
APR techniques.

In summary, this paper makes the following contributions:
\begin{itemize}
	\item We propose a novel variable-based fault localization technique, named \tool{}, which identifies fault-relevant variables via an enhanced decision tree model.
	\item We conduct an extensive study on the widely-used Defects4J benchmark in two distinct application scenarios. The results demonstrate the effectiveness of our approach by comparing it with existing state-of-the-art approaches.
	\item We provide a new way of thinking for improving APR techniques -- providing finer-grained fault localization results to refine the patch space of APR tools.
	\item We have published all our experimental results and implementation to facilitate future research for replication and comparison.
\end{itemize}


\section{Motivating Example}
\label{sec:motivate}
In this section, we will motivate our approach with a running example.
	Listing~\ref{lst:lang-27} presents the patch code of Lang-27 in the widely used Defects4J benchmark~\cite{just2014defects4j}, where the lines leading by ``+'' denote code to be added while ``-'' to be deleted. 
	
	\newcommand{\hlc}[2]{{\setlength\fboxsep{0pt}\hspace{-3pt}\colorbox{#1} 
		{\begin{minipage}{\dimexpr\columnwidth-1\fboxsep+0pt\relax}
				\codeIn{\strut\hspace{3pt}#2}
			\end{minipage}}}}

\begin{lstlisting}[style=java, numbers=none,label=lst:lang-27,caption=Patch of Lang-27.]
<@\lin{452}@> Number createNumber(String str) throws Exception {
<@\lin{}@>  ...
<@\hlc{gray!15}{\lin{473}\qquad int decPos = str.indexOf('.');}@>
<@\hlc{gray!15}{\lin{474}\qquad int expPos = str.indexOf('e') + str.indexOf('E') +1;}@>
<@\lin{475}@>
<@\hlc{gray!15}{\lin{476}\qquad if (decPos > -1) \{ }@>
<@\lin{477}@>
<@\lin{478}{\qquad\qquad if (expPos > -1) \{ }@>
<@\hlc{red!15}{\lin{479}\,- \qquad\qquad   if (expPos < decPos) \{ }@>
<@\hlc{green!15}{\qquad  + \qquad\qquad   if (expPos < decPos || expPos > str.length())\{ }@>
<@\lin{480}@>            throw new NumberFormatException(str + " is not a valid number.");
<@\lin{481}@>         }
<@\lin{482}@>         dec = str.substring(decPos + 1, expPos);
<@\lin{483}{\qquad\qquad     \} else \{ }@>
<@\lin{484}@>         dec = str.substring(decPos + 1);
<@\lin{485}{\qquad\qquad     \} @>
<@\lin{486}{\qquad\qquad     mant = str.substring(0, decPos); @>
<@\lin{487}{\qquad \} else \{ } @>
<@\hlc{gray!15}{\lin{488}\qquad\qquad if (expPos > -1) \{ }@>
<@\hlc{green!15}{\qquad  + \qquad\qquad   if (expPos > str.length()) \{ }@>
<@\hlc{green!15}{\qquad  + \qquad\qquad\qquad throw new NumberFormatException(str + " is not a valid number."); }@>
<@\hlc{green!15}{\qquad  + \qquad\qquad   \} }@>
<@\hlc{gray!15}{\lin{489}\qquad\qquad\quad\   mant = str.substring(0, expPos);}@>
<@\lin{}@>  ...
\end{lstlisting}


	In this example, when providing an input string \codeIn{str}, the method \codeIn{createNumber(*)}
	will transform it into a \codeIn{java.lang.Number} object, e.g., transforming ``10'' into an \codeIn{Integer} of 10.
	In this process, the method will automatically check the validity of the input and then decide which type of number should be created. For example, when the input string contains the character ``\codeIn{e}'' (or ``\codeIn{E}''), an exponential number is always expected. However, due to the faulty code, when taking the illegal input ``\codeIn{1eE}'', a \codeIn{StringIndexOutOfBoundsException} was incurred at line 489 (line 479 can be triggered by other inputs), while actually a \codeIn{NumberFormatException} was expected (see Listing~\ref{lst:lang-27}). The reason is that the method failed to check the validity of the input when multiple ``\codeIn{e/E}''s exist.

\begin{figure}[]
		\begin{minipage}{.58\columnwidth}
			\resizebox{\columnwidth}{!}{%
				\begin{threeparttable}
				\setlength\tabcolsep{2pt}
				\begin{tabular}{c|cccc}
					\toprule
					\textbf{Test}        & \textbf{\codeIn{str}}     & \textbf{\codeIn{expPos}} & \textbf{\codeIn{str}.\codeIn{length()}} & \textbf{Result} \\ \midrule
					$t_1$     &     ``1l''        & -1 & 2  & PASS \\
					$t_2$     &      ``1111 ''      & -1 & 5  & PASS \\
					$t_3$ &  ``-1.1E200'' & 4  & 8  & PASS \\
					$t_4$ & ``1eE''  & 4                               & 3         & {\color[HTML]{FE0000} FAIL}              \\ \bottomrule
			\end{tabular}%
		  \begin{tablenotes}
			    \small
			    \item Test samples of Lang-27 and partial variable values in the faulty method when running the corresponding test.
			\end{tablenotes}
		\end{threeparttable}
		}
	\end{minipage}\hfill
	\begin{minipage}{0.4\columnwidth}
		\centerline{\includegraphics[width=0.99\columnwidth]{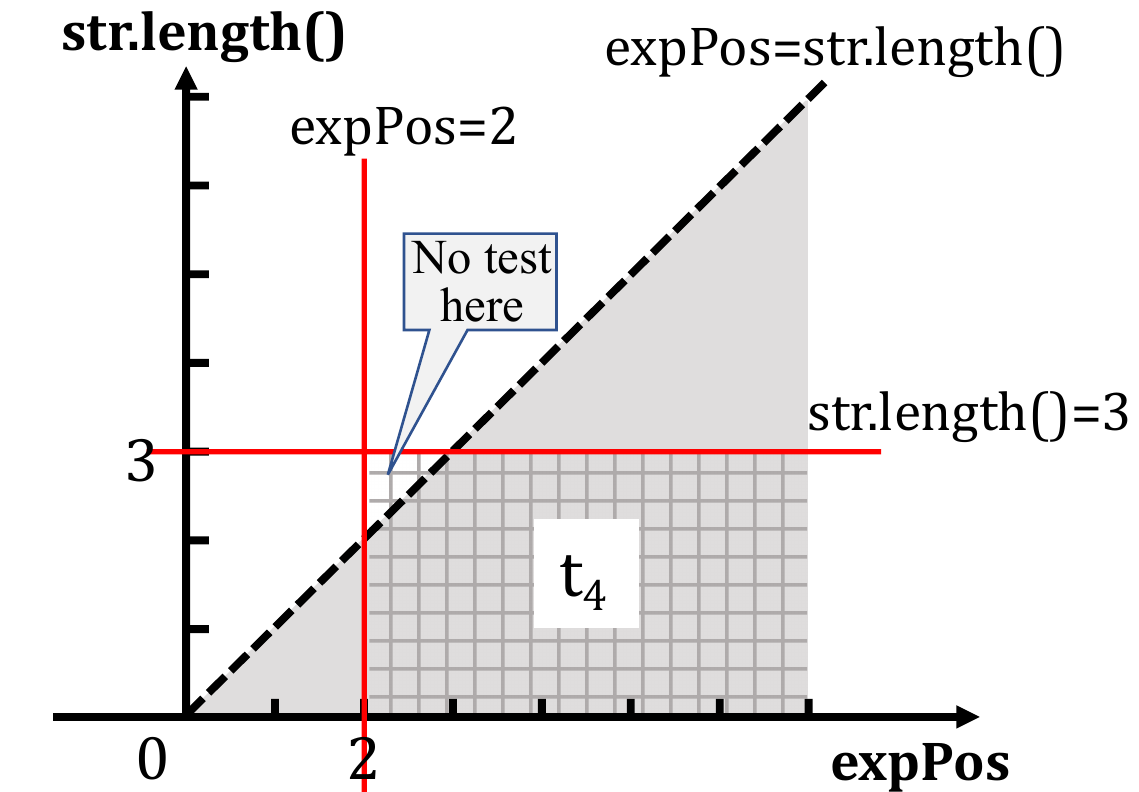}}
	\end{minipage}
	\caption{Visualization of variable constraint estimation.}
	\label{fig:visualize}
\end{figure}

	To locate the root cause of the failure, existing approaches typically return a ranked list of suspicious code lines (or methods), such as the widely-used coverage-based fault localization techniques~\cite{abreu2007accuracy,zou2018empirical}. However, existing approaches can hardly locate the accurate faulty code in this example.
	In fact, 
	even providing the genuine faulty code line, there is still a large search space (i.e., any syntax-valid expressions)
	to repair the bug due to the coarse-grained fault localization results, where incorrect patches may also be easily produced.
	On the contrary, 
	if the finer-grained fault-relevant variables \codeIn{expPos} and \codeIn{str.length()} were known, the patch space would be significantly reduced and thus incorrect patches would also be effectively avoided. 
	
	
	However, accurately identifying the fault-relevant variables is indeed challenging since the variable values can be diverse in different test runs (see the left table in Figure~\ref{fig:visualize}). 
	Besides, it is also common that we are required to capture the complex constraints among multiple variables for isolating failed from passed runs and understanding the root cause, e.g., \codeIn{expPos>str.length()}.
	However, checking all possible variable combinations is indeed time-consuming and even impossible in practice.
	%
	%
	%
	Targeting this challenge, we hereby propose a novel variable-level fault localization technique based on the \textit{decision tree model}. The basic intuition of our approach is that complex variable constraints can be estimated (or even constructed) by combining multiple primitive constraints, where only one variable is used in each individual constraint. The reason is that each primitive constraint can discriminate the failed test run from at least a subset of the passed runs and their combinations may approximate the desired complex constraint. For example, the primitive constraints \codeIn{expPos>=2} and \codeIn{str.length()<4} can discriminate $t_4$ from $\{t_1,t_2\}$ and $\{t_2,t_3\}$ respectively, and their combination can estimate \codeIn{expPos>str.length()} in the running example as shown in Figure~\ref{fig:visualize} (right-side figure). In the figure, the \textbf{shaded area} denotes the constraint of \codeIn{expPos>str.length()}, while the \textbf{gridded area} represents the two primitive constraints. Therefore, the failed ($t_4$) and passed test cases ($t_1,t_2$ and $t_3$) can also be distinguished by the two primitive constraints. In this way, the variables used, e.g., \codeIn{expPos}, in those primitive constraints have large possibility to be the indicator of the test failure since it has the ability to isolate the failed tests from the passed ones, and thus are potentially the fault-relevant variables (defined in Section~\ref{sec:measure}).
	However, since there are usually many available variables that may produce a large number of primitive constraints, how to combine them and correctly locate the desired fault-relevant variables is still \textit{non-trivial}. According to the characteristics of this task, we propose an enhanced decision tree model to aid the variable identification and constraint building process since the decision tree in nature performs a similar process to our task, i.e., using multiple primitive constraints (branch conditions) to estimate complex constraints for classification. We will introduce more details in Section~\ref{sec:approach} by taking this bug as the running example.

\section{Framework}
\label{sec:approach}

\newcommand{\step}[1]{\textcircled{#1}}

\begin{figure*}[t]
	\centering
	\includegraphics[width=0.99\textwidth]{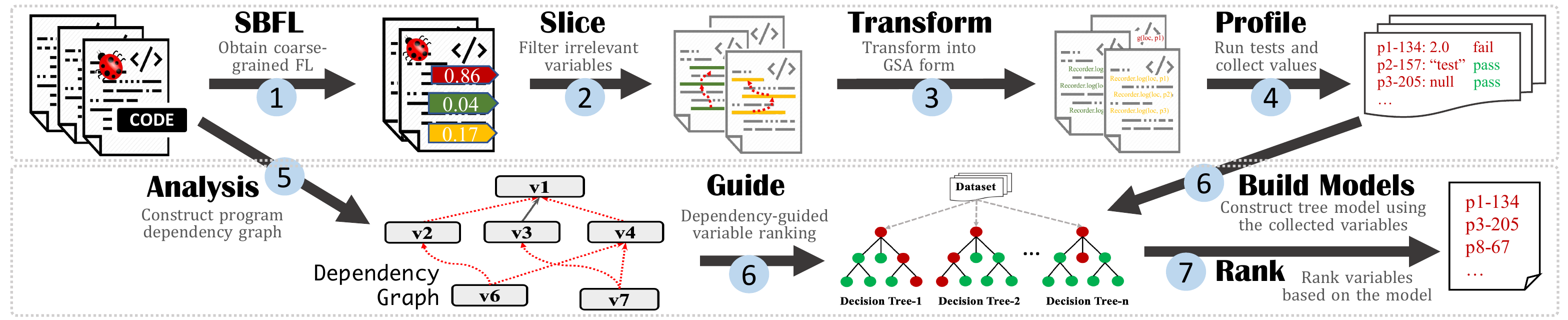}
	\caption{Overview of our approach \tool{}}
	\label{fig:overview}
\end{figure*}

This section introduces the details of our approach. As aforementioned, the basic idea of our approach is to use variables to build constraints for distinguishing failed and passed runs, where the variables that have higher discrimination ability have larger possibilities to be the fault-relevant variables. Figure~\ref{fig:overview} shows the overview of our approach (named \tool{}). In general, it consists of two stages. When given a program with at least one test case failed on it, \tool{} first \textit{collects the values of a set of variables} in both failed and passed test runs at some program checkpoints. Then, it \textit{builds decision tree models} using those collected variables to distinguish failed and passed test runs, after which it identifies the fault-relevant variables from those used for constructing the branch conditions (i.e., constraints) in the models since they exhibit the ability to discriminate failed and passed tests.

However, it is hard and even impossible to examine the complete space of all program variables since it is usually huge, especially for large-scale programs, which may involve tens of thousands of variables. To overcome this challenge, \tool{} combines existing lightweight method-level fault localization and adopts program slicing technique to identify a subset of covered statements for inspection, which can improve the efficiency and scalability of our approach.
Specifically, in the current implementation of \tool{}, we utilize the widely-used spectrum-based fault localization (SBFL) technique to locate a list of the most suspicious methods.
Particularly, we adopt the implementation published by Jiang et al.~\cite{jiang2019combining}.
Please note that our approach is independent of this localization process, and it can be easily replaced by other methods as long as the output is an ordered list of suspicious faulty methods, such as the latest deep-learning-based techniques~\cite{lou2021boosting,li2021fault}, which can produce much better results than SBFL and potentially can further improve the performance of \tool{}. 


\subsection{Dynamic Program Slicing}
\label{sec:slice}

By using the coarse-grained fault localization techniques, we can obtain an ordered list of methods that are most likely to contain bugs. In this way, we can just focus on the variables used in these methods. However, it is intuitive that not all statements and variables in these methods affect the output of the failing tests.
In order to further reduce the search space of candidate variables for inspection and increase accuracy, \tool{} leverages dynamic program slicing techniques~\cite{dynamic-slicing} to filter out statements that are indeed irrelevant.

Specifically, when given a slicing criterion and a certain test input, \tool{} performs an intra-procedure slicing process based on the data and control dependency relations along the execution trace backwardly. Although less accurate compared with the inter-procedure slicing, the intra-procedure slicing is much more efficient without the need of heavy inter-procedure analysis. As a consequence, the slicing process in \tool{} will be not affected by the scale of programs under debugging but only affected by the size of a single method. Regarding the slicing criterion, we pick \textit{the line of code that was lastly executed by the failed test in the method} because it is usually the location of failures or the indicator of finishing the complete functionality of the method and may produce variables affecting the subsequent program execution (e.g., \codeIn{return} statements in many cases). For instance, recall the example shown in Listing~\ref{lst:lang-27}, the failed test run crashed at line 489 (lastly executed), which directly depends on the fault-relevant variables \codeIn{str} and \codeIn{expPos}, and thus they will be included in the slicing while the variable \codeIn{mant} in line 486 will be filtered out.
In this way, a subset of statements will be identified for further checking, highlighted in the gray color \crule[gray!50] in Listing~\ref{lst:lang-27} (Lines 473-476,488,489), while the other statements and associated variables will be ignored.

In our evaluation, we will also conduct an experiment to discuss the impact of the slicing process on the effectiveness of our approach in Section~\ref{sec:result}.

%
%
%

\subsection{Program Transformation and Profiling}
\label{sec:transform}

By program slicing, a subset of statements that are most likely to be the root cause of the test failure can be obtained. Next, \tool{} will collect the variable values in those statements during the running of test cases by automatically instrumenting output statements to the source code. 

Particularly, in order to tackle programs of any forms, \tool{} further incorporates a program transformation process that can transform source code into a GSA form~\cite{kucuk2021improving}, where compound expressions will be implicitly decomposed into TAC (Three Address Code) format. For example, the expression \codeIn{(a>b\&\&c>d)} will be transformed into \codeIn{(v=((v$_1$=(a>b))\&\&(v$_2$=(c>d))))} by inserting corresponding temporary variable declarations on demand (i.e., \codeIn{v}, \codeIn{v$_1$} and \codeIn{v$_2$}). In this way, the intermediate computation results of compound expressions can also be collected through these temporary variables, such as the result of \codeIn{a>b}.
Specifically, \tool{} transforms expressions in three types of code structures, i.e., \textit{conditional expressions}, \textit{return expressions} and \textit{arguments of method calls}. The reason is that conditional expressions are error-prone in practice and many bugs are caused by incorrect sub-conditions~\cite{ISSTA18-SimFix,Wen2018ContextAwarePG,liu2019tbar}, while the expressions in the latter two types take the responsibility of value transmission and thus may potentially spread faulty variable values to a broader range outside the method. As a result, checking the values of these expressions is indeed necessary for locating the root causes of program failures.

After program transformation, \tool{} can only focus on the variables (including temporary variables of expressions) used by the statements in the slicing.
%
%
Specifically, apart from the concrete values exhibited by the (temporary) variables used in the program, we have also defined several common predicates that may be highly related to test failures, such as checking whether an object is \codeIn{null}. We have summarized the details of values collected for different types of variables in Table~\ref{tab:univ-compa}. 
Based on this definition, the values that will be collected at line 489 in Listing~\ref{lst:lang-27} include not only the primitive variable values of \codeIn{expPos} and \codeIn{mant}, but also the predicate values of \codeIn{str == null} and \codeIn{str.length()}.

\begin{table}[tb]
	\centering
	\caption{A description for variables considered by \tool{}}
	\label{tab:univ-compa}
	\resizebox{\columnwidth}{!}{%
	\begin{tabular}{ccc}
		\toprule
		\textbf{Type} & \textbf{Target Value} & \textbf{Description}\\ \midrule
		Primitive & Actual Value & Primitive value or ASCII code for \codeIn{char} \\ \hline
		\multirow{5}{*}{Object} & Null Check & True if the variable is \codeIn{null}, false otherwise\\ 
		& Type Check & The value type of the variable. e.g., \codeIn{String} \\ 
		& Fields & Unfold the variable and output field values \\ 
		& Size/length  & Access \codeIn{size()}/\codeIn{length()}/\codeIn{length} (if has) \\
		& Elements & Primitive element values in collections \\
		\bottomrule
	\end{tabular}
	}
\end{table}

To collect the above variable values, we have implemented a simple value profiling process in \tool{}, which can automatically parse the type of fed variables and insert output statements by using the Eclipse Java Development Toolkit (https://www.eclipse.org/jdt/) for recording the corresponding values defined in Table~\ref{tab:univ-compa} during the running of test cases.

\subsection{Tree Model Construction}
\label{sec:model}

As aforementioned, the basic idea of our approach is using variables to construct (multiple) primitive constraints and their combinations to distinguish passed and failed test runs, where the variables that have higher discrimination ability may have larger possibility to be fault-relevant.
Based on this, we propose a novel fault-relevant variable identification technique by leveraging the decision tree model, which has been well studied to be effective in many applications~\cite{deng2019feature,gupta2020novel,xiong2022l2s,tizpaz2018differential}. Particularly, this model is suitable for our application in two aspects. (1) Our application scenario actually can be viewed as a binary classification task, where the labels are ``PASS'' and ``FAIL'', representing the testing results of test cases. (2) The decision tree model has good interpretability, where the branch conditions in the model explain how a given input is classified to the particular class. The conditions in the same tree path can be combined to form a more complex constraint that is only satisfied by the data belonging to the corresponding leaf node in the tree.
That is, why an input is classified to the corresponding class is traceable. Recall that our ultimate target is to identify the variables that can discriminate failed and passed tests, the interpretability and traceability properties of the model satisfy our requirements.

\begin{algorithm}[t]
	\scriptsize
	\caption{Variable Prioritization for Selection}
	\label{alg:prioritize}
	\SetKwInput{KwInput}{Input}                
	\SetKwInput{KwOutput}{Output}              
	\DontPrintSemicolon
	
	\KwInput{\textit{varList:} \textcolor{gray}{a list of variables to be ranked.}
		\textit{data:} \textcolor{gray}{a set of program states for each test case.}
		\textit{graph:} \textcolor{gray}{program dependency graph.}}
	\KwOutput{\textit{varList:} \textcolor{gray}{an ordered list of variables}}
	
	\SetKwFunction{FRank}{prioritizeVars}
	
	\SetKwProg{Fn}{Function}{:}{}
	\Fn{\FRank{\textit{varList, data, graph}}}{
		\ForEach{var in varList}{
			\textit{var.score} $\leftarrow$ \textit{gainRatio(var)} + \textit{correlation(var, data.labels)}\;
		    \textit{dependency} $\leftarrow$ \textit{depScore(var, graph, varList)}\;	\textit{var.score} $\leftarrow$ \textit{var.score} $\times$ \textit{dependency}
		}
		\ForEach{var in varList}{
		    \ForEach{v in var.getEqualVars(graph)}{
		        \textit{var} $\leftarrow$ \textit{aggregate(var, v)}\;
		        \If{v.score $>$ var.score}{
					\textit{var.score} $\leftarrow$ \textit{v.score}\;
				}
		    }
		}
		\KwRet \textit{varList.sort()} \tcp*{descending order by $score$}
	}
\end{algorithm}

Next, we will introduce the details of our tree model construction process in \tool{}. In general, it includes two sub-processes, named \textit{Enhanced Variable Selection} and \textit{Tree Model Building}. The former takes the responsibility to select proper variables for branch condition building, while the latter then uses the selected variables to construct concrete conditions and divides test runs into different groups. For each group, the same process will proceed until the tests in all groups cannot be further divided, where a decision tree is built.


\subsubsection{Enhanced Variable Selection}
Unlike the features used in traditional classification problems, variables collected by \tool{} naturally have clear and strong correlations, i.e., control dependency and data dependency, which reflect the impacts of different variables to the execution results. For example, in the patch code shown in Listing~\ref{lst:lang-27},
the crashed line 489 depends on the variable \codeIn{expPos} defined in line 474, which further depends on the input argument \codeIn{str}. In other words, though the program crashed due to the incorrect value of \codeIn{expPos}, the input \codeIn{str} may also be the root cause of the failure in practice.
However, the general variable selection algorithm in decision tree models do not consider such dependency information, and may significantly affect the overall effectiveness of fault-relevant variable localization since it may cause the irrelevant variables located and decrease the fault localization precision (refer to Section~\ref{sec:eval}). To overcome this limitation, we propose an \textit{enhanced variable selection strategy} depending on a novel variable prioritization algorithm which takes the program dependency factor into consideration.

 Intuitively, when a variable is depended on by more other variables, its value will have higher possibility to affect the final execution results in different execution paths, and thus potentially affect more test cases. However, we observe that usually a small number of test cases, e.g., one or two, will be affected and failed in real-world buggy programs. In other words, the faulty variables tend to affect test cases in a small scale. Therefore, we introduce a \textit{dependency penalty} to incorporate such an observation through static analysis.
 That is, a variable depended on by more other variables will rank lower, i.e., less likely to be faulty.
 Formula~\ref{eq:dep} defines the computation of the penalty for variable $v$ when providing the dependency graph $g$ and a list of interested variables $l$ in $g$.
\begin{equation}\label{eq:dep}\small
	\begin{aligned}
		depScore(v,g,l)=\depfactor{}^{|S|}\\ 
		s.t.~S=\{x|x\in l \wedge g\vdash x\hookrightarrow v\}
	\end{aligned}
\end{equation}
In the formula, we use $g\vdash x\hookrightarrow v$ to represent that variable $x$ depends on variable $v$ according to $g$, i.e., the node of $v$ in graph $g$ is reachable from that of $x$. $\depfactor{}\in(0, 1.0]$ is a constant penalty factor, indicating how much the dependency affects the importance of variables.

Based on this definition, we present our variable prioritization algorithm in Algorithm~\ref{alg:prioritize}. Specifically, for each variable $var$,
its priority is determined by three parts (lines 3-5). The \textit{dependency penalty} has been defined in Formula~\ref{eq:dep}, while
the function of \textit{correlation(*)} returns the general \textit{Pearson correlation coefficient}~\cite{benesty2009pearson} between variables and the testing results.
Finally, the \textit{gainRatio(var)} is a builtin function in the C4.5 decision tree model~\cite{quinlan2014c4} for computing how much confidence can be gained by choosing the variable \textit{var} to distinguish the given data.
In summary, a variable that has a smaller impact to the program semantics (i.e., larger \textit{depScore(*)}), a larger correlation to the test results, and more confidence to be the discriminator, will gain higher priority.

After computation, each variable will be assigned a priority score (refer to lines 2-6 in Algorithm~\ref{alg:prioritize}). Next, we aggregate the equivalent variables appearing at different
locations (i.e., no reassignment between them) into one as the representative by removing the others according to the dependency relation (lines 8-9), and the score of the representative variable will be the maximal one of all its equivalent variables (lines 10-11). The reason is that they are always having the same value in a run, which may cause duplicate selection of the same variable. For example, the variables of \codeIn{expPos} appearing at lines 474, 488 and 489 in Listing~\ref{lst:lang-27} are equivalent, then two of them will be removed in the results returned by Algorithm~\ref{alg:prioritize}. Finally, variable with larger score will have higher priority to be selected during the tree model building process.

%
%
%
%
%
%

\begin{algorithm}[tb]
	\scriptsize
	\caption{Tree Model Building}
	\label{alg:model}
	\SetKwInput{KwInput}{Input}                
	\SetKwInput{KwOutput}{Output}              
	\DontPrintSemicolon
	
	\KwInput{\textit{data:} \textcolor{gray}{a set of program states for each test case.}\;
		 \qquad \textit{graph:} \textcolor{gray}{program dependency graph.}}
	\KwOutput{\textit{trees:} \textcolor{gray}{a set of decision trees.}}
	
	\SetKwFunction{FMain}{buildModel}
	\SetKwFunction{FBuild}{buildTree}
	
	\SetKwProg{Fn}{Function}{:}{}
	\Fn{\FMain{\textit{data, graph}}}{
		\textit{trees} $\leftarrow \emptyset$\;
		\textit{varSet} $\leftarrow$ \{\textit{var} | \textit{var} is recorded in \textit{data} \}\;
		\While{varSet is not empty}{
			\tcc{build multiple trees with all variables}
			\textit{tree $\leftarrow$ \FBuild{data, toList(varSet), graph}}\;
			\If{tree is not a leaf node}{
				\textit{trees $\leftarrow$ trees $\cup$ tree}
			}
			\textit{varSet $\leftarrow$ varSet $\setminus$} \{\textit{var} | \textit{var} is used by \textit{tree}\}
		}
	\KwRet \textit{trees}
	}
	\Fn{\FBuild{\textit{data, varList, graph}}}{
		\textit{tree} $\leftarrow$ \textit{leafNode(data)} \;
		\tcc{size($data$)>2 $\wedge$ $data$ include different labels}
		\If{data can be classified}{
			\tcc{prioritize different attributes}
			\textit{varList} $\leftarrow$ \FRank{\textit{varList, data, graph}}\;
			\textit{var} $\leftarrow$ \textit{varList.first} \tcp*{higher priority first}
			\textit{cond} $\leftarrow$ \textit{calculateCondition(data, var)}\;
			\tcc{divide $data$ into groups based on $cond$}
			\textit{groups} $\leftarrow$ \textit{divide(data, var, cond)} \;
			\textit{tree $\leftarrow$ internalNode(data)} \tcp*{root node of subtree}
			\ForEach{g in groups}{
				\textit{tree.children.add}(\FBuild{\textit{g, varList}})
			}
		}
		\KwRet \textit{tree}
	}
\end{algorithm}

\subsubsection{Tree Model Building}
According to the above variable selection strategy, we present the details of our model building process, which is shown in Algorithm~\ref{alg:model}. When providing the values of a set of variables per test case (i.e., \textit{data}) and the dependency graph of the program, the tree model building process (i.e., \textit{buildModel(*)}) is iteratively proceeded. That is, \tool{} each time chooses a subset of variables to construct a tree model (line 5) until using up all variables (line 4). As shown in line 9, each variable can be used in no more than one decision tree to avoid duplication and guarantee this process always terminates. In other words, the output of the model building process is a set of decision trees, each of which can independently isolate the failed test runs from the passed ones by using a subset of variables. In this way, all variables will have the possibility to be located since the fault-relevant variables can be multiple.
Particularly, in each iteration, the tree model is recursively constructed from the top down using the provided variables by invoking the function of \textit{buildTree(*)}. Specifically, each time the variable with the highest priority (i.e., \textit{varList.first}) will be selected to construct a predicate for dividing the given \textit{data} into different groups (lines 15-18). If the selected variable $var$ is nominal, the predicate will be a switch-case-like multi-way condition, while if numeric, a binary predicate, such as ``$\ge$'' and ``$<$''  will be generated. According to the predicate, \textit{data} will be divided into different groups. \tool{} recursively performs the above construction process (lines 20-22) for each group until the input data do not require further discrimination (line 14).

\begin{figure}[]
	\centering
	\includegraphics[width=0.98\columnwidth]{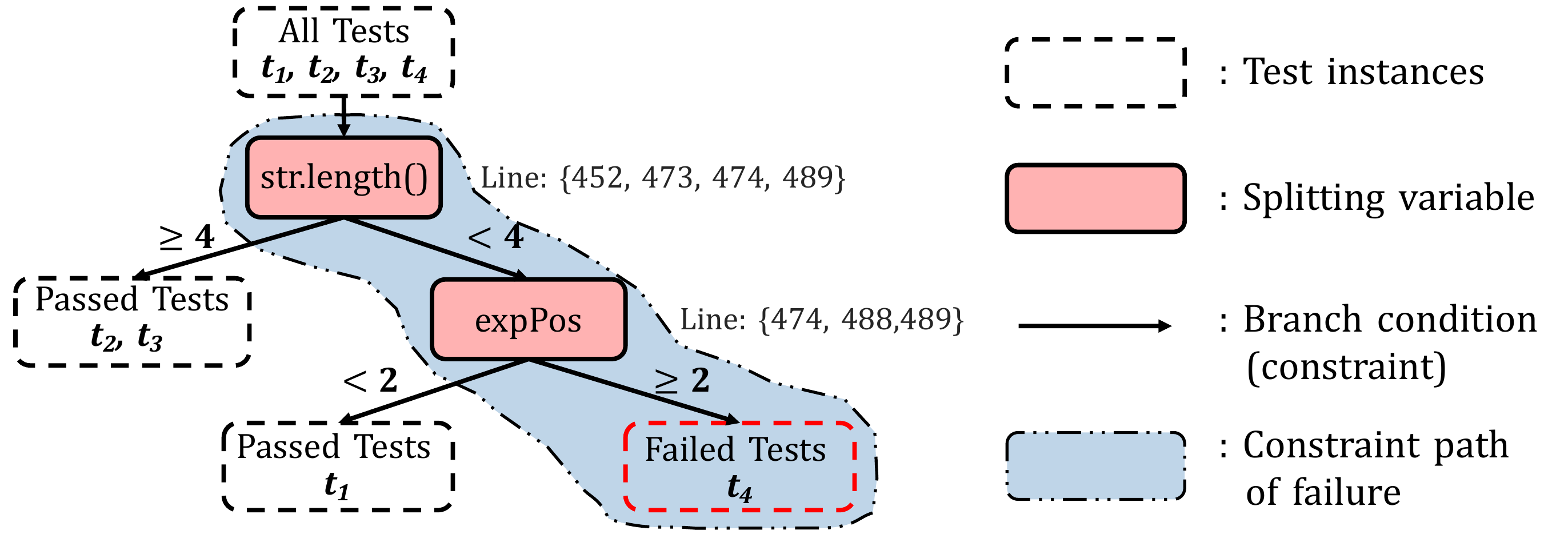}
	\caption{A sketch of tree construction for Listing~\ref{lst:lang-27}}
	\label{fig:tree-pic}
\end{figure}

Up to now, when providing the required data, tree models can be constructed according to Algorithms~\ref{alg:prioritize} and~\ref{alg:model}. For example, recall the example shown in Listing~\ref{lst:lang-27}, according to Algorithm~\ref{alg:prioritize}, the temporary variable representing \codeIn{str.length()} will receive the highest priority, and thus will be first selected for building the branch condition as shown in Figure~\ref{fig:tree-pic}. Specifically, according to its values in different test runs (see Figure~\ref{fig:visualize}), a branch condition \codeIn{str.length()<4} will be constructed and divide tests into different groups\footnote{Please note that the constant value ``4'' is automatically computed by the default builtin function of decision tree model in Weka (https://www.weka.io).}, i.e., $\{t_2, t_3\}$ and $\{t_1, t_4\}$. Recursively, in the second round variable \codeIn{expPos} will be selected and further divide the test set $\{t_1, t_4\}$ into $\{t_1\}$ and $\{t_4\}$. By now, the failed test run ($t_4$) is completely isolated from the passed runs. From Figure~\ref{fig:tree-pic} we can also see that the constraints only satisfied by the failed test runs are highly related to the root cause of the failure.

In particular, to improve the scalability and efficiency, \tool{} builds tree models for different methods independently. That is, \tool{} each time takes the profiled variable data and the intra-procedure dependency graph within a single method as the input and outputs a set of constructed models, based on which it identifies the most fault-relevant variables by a ranking strategy (to be presented in Section~\ref{sec:rank}). 


\subsection{Variable Ranking}
\label{sec:rank}
According to the previous sections, when providing a buggy program, \tool{} can construct a set of tree models for each candidate faulty method using the associated variables. In this section, we further introduce the variable ranking strategy, which provides a protocol to rank variables in different models of different methods and obtain the list of the most suspicious variables that are fault-relevant. Please note that this ranking strategy is different from the variable prioritization process shown in Algorithm~\ref{alg:prioritize}, where the latter aims to make the most suspicious variables be chosen for model building by estimating their capability of discriminating between failed and passed tests, while the former ranking strategy to be introduced in this section is to assign a global suspicious score to each chosen variable according to the built models.

Specifically, we define a decision tree as a tuple of $M=(t, p, D, C)$,
where $t$ denotes the root node of the tree, $p$ denotes the predicate associated with the node $t$, and $D$ is a set of data (including tests and corresponding variable records) associated with the node $t$. Finally, $C$ is the set of subtrees of $t$. Then, when providing the tree $M$ built for a particular method, we define the posterior discriminative score of variable $v$ used in the predicate $p$ by Formula~\ref{eq:pscore}.
\begin{equation}\label{eq:pscore}\small
	DS(v) = \frac{(1-Gini(p))\times\sqrt{|D|}}{\mathrm{\textit{failNodeDist}}}+{depScore(v, g, l)}
\end{equation}
where $|D|$ denotes the number of test cases in $D$, we use its square root to shrink the discrepancy of test numbers since it can range from several to hundreds.
$Gini(p)$ represents the general \textit{Gini index}~\cite{breiman1984classification} of predicate $p$, denoting the \textit{impurity} of the tree rooted $t$.
\textit{failNodeDist} denotes the length of the tree path from the root node $t$ to the leaf node containing the failed tests in $M$. 
The smaller the length is, the more specific to the failed test the variable will be, and thus will be more fault-relevant.
The second part $depScore(v,g,l)$ is the \textit{dependency penalty} of variable $v$ defined by Formula~\ref{eq:dep}. 
To sum up, the score of variable $v$ is determined by the discrimination ability of the variable in the decision tree (the first part), and its impact on the program semantics (the second part).


Then, the global ranking score of variable $v$ from method $m_v$ is computed by Formula~\ref{eq:rscore}, where $methodScore(m_v)$ denotes the suspiciousness of method $m_v$ computed in the first step of \tool{}, i.e., the method-level fault localization. In particular, we use the quadratic value of the suspiciousness to weaken its impact on the final rank and thus strengthen the importance of the variable discrimination ability (i.e., $DS(v)$). The bigger the $FS(v)$ is, the higher the variable $v$ will rank.
\begin{equation}\label{eq:rscore}\small
	FS(v) = {DS(v)}\times{methodScore(m_v)^{2}}
\end{equation}
According to this ranking strategy, the fault-relevant variable \codeIn{expPos} at lines $\{474, 488, 489\}$ was successfully ranked at the Top-1 position.
As shown in Figure~\ref{fig:tree-pic}, the built constraints related to the failure can indeed estimate the desired complex constraints as presented in Figure~\ref{fig:visualize}. 
\section{Experiment Setup}
\label{sec:eval}

To evaluate the effectiveness of our approach, we have implemented it in a tool named \tool{}, and conducted an extensive study by comparing it with state-the-the-art fault localization approaches. Besides, to investigate whether our finer-grained fault localization results can further improve the effectiveness of downstream APR approaches, we also adapted our approach to the application of patch filtering. Specifically, we address the following research questions in our evaluation.

\begin{itemize}
	\item \textbf{RQ1:} How effective is \tool{} for identifying fault-relevant variables in real-world programs?
	\item \textbf{RQ2:} Does each component contribute to the effectiveness of \tool{}?
	\item \textbf{RQ3:} Can \tool{} help to improve the results of automatic program repair?
\end{itemize}

\subsection{Subjects}
\label{sec:subject}
In the evaluation of fault localization (RQ1\&RQ2), we adopt the Defects4J (version 2.0) benchmark~\cite{just2014defects4j}, which is widely-used in previous studies~\cite{jiang2019combining,zou2018empirical,Michael:SBFL,liu2019tbar,ISSTA18-SimFix,Wen2018ContextAwarePG}.
Specifically, we conducted our experiment on a subset of the benchmark according to the following two constraints. First, the genuine faulty method can be located within the Top-10 returned results by the method-level fault localization in the first step of \tool{} as shown in Figure~\ref{fig:overview}. The reason is that the state-of-the-art approach can correctly locate more than 80\% bugs
in Top-10~\cite{lou2021boosting}. Therefore, targeting this portion of bugs can be significant for practical use and also reduce the overhead of variable profiling. 
Second, the faulty method has to be covered by at least three (at least one failed and one passed) test cases so as to the decision tree model in \tool{} can work normally. The details of the subjects used in our experiment are listed in Table~\ref{tab:fl-subject}. Regarding the patch filtering application (RQ3), we adopt the dataset constructed by Xiong et al.~\cite{xiong-icse18} and use all the patches for bugs included in Table~\ref{tab:fl-subject}.

\begin{table}[t]
	\centering
	\caption{Subjects for fault localization}
	\label{tab:fl-subject}
	\resizebox{\columnwidth}{!}{%
		\begin{tabular}{lc|lc|lc}
			\toprule
			\textbf{Project}         &  \textbf{\#Bugs}  & \textbf{Project}         &  \textbf{\#Bugs}   & \textbf{Project}         &  \textbf{\#Bugs}  \\
			\midrule
			Math  & 23 &  JxPath & 8 & Compress & 9 \\
			Chart & 12 & Cli & 8 &  JacksonXml (JXml) & 4 \\
			Lang  & 23 & Gson & 7 & JacksonCore (JCore) & 16 \\
			Time & 12 & Csv & 5 & 
			Mockito & 21\\
			Codec & 10 & Jsop & 18 & JacksonDatabind (JDatabind) & 28 \\
			\midrule
			\multicolumn{2}{l}{\textbf{TOTAL}} & \multicolumn{4}{|c}{204} \\
			\bottomrule
		\end{tabular}%
	}
\end{table}

\subsection{Baseline and Configuration}
\label{sec:baseline}
In the experiment of fault localization, following the latest research~\cite{kucuk2021improving}, we compare the effectiveness of our approach with five state-of-the-art variable-level fault localization techniques: \textbf{UniVal}~\cite{kucuk2021improving}, the latest approach that 
uses causal inference and machine learning to integrate information about both predicate outcomes and variable values to estimate the effects of variables to test failures;
NUMFL~\cite{bai2015numfl} (specifically the two variants \textbf{NUMFL-QRM} and \textbf{NUMFL-DLRM}), locating variables by combining causal and statistical analyses to characterize the causal effects of individual numerical expressions on output errors;
\textbf{ESP}~\cite{core2021statistical}, locating variables via measuring the difference between an assigned variable in the failed run and its average value in all test runs;
and \textbf{Baah2010}~\cite{Baah2010causal}, locating variables by using a linear regression model to measure the
\textit{confounding bia} among variables.
Specifically, we adopt their open-source implementation published by K\"{u}\c{c}\"{u}k et al.~\cite{kucuk2021improving} to perform the experiment. Besides, we also adapt two representative and most widely-used spectrum-based fault localization techniques to work at variable level, i.e., \textbf{Ochiai}~\cite{Ochiai} and \textbf{DStar}~\cite{DStar}, which were proved to perform well~\cite{Michael:SBFL,Lucia2014extended}.

Regarding the configurations of \tool{}, we set the Top-10 most suspicious methods as the interested ones as explained in Section~\ref{sec:subject}, and set the default value of \depfactor{} as 0.8 for computing the \textit{dependency penalty} in Formula~\ref{eq:dep}, whose impact will be further investigated in our evaluation. In addition, to evaluate the effectiveness of each component in our approach, we also create a set of variants of \tool{}.
\begin{description}
    \item[\variant{slice}]: removes the dynamic program slicing component in \tool{} and considers variables in all statements covered by the failed tests within the interested methods.
    \item [\variant{tree}]: removes the tree model in \tool{}. As a result, the variables are basically ranked according to the \textit{dependency penalty} and the method suspicious score.
    \item [\variant{dep}]: removes the \textit{dependency penalty} used for variable ranking from both model building and variable ranking processes, while keeps the others unchanged.
    \item [\variant{ms}]: removes the method score in the final variable ranking process of \tool{}, i.e., $FS(v)=DS(v)$.
\end{description}

Particularly, since our approach is not designed as a standalone patch filtering tool, we further adapt it to this scenario. Specifically, we perform this process by simply using the located fault-relevant variables to filter patches directly. If no fault-relevant variable is involved in the patch, i.e., not modified or inserted, the patch will be filtered, otherwise regarded as correct.
In this study, we compare the results of our approach with PATCH-SIM~\cite{xiong-icse18}, the state-of-the-art patch filtering technique based on generating new test cases.

To ease the replication of our experimental results and promote future studies in this research area, we have published all our experimental data and the implementation of \tool{} at \datalink{}.

\subsection{Measurement}
\label{sec:measure}

Although several variable-level fault localization techniques have been proposed as introduced in the Introduction, there is still no clear definition of fault-relevant variables, the ground truth employed by different studies may also be diverse. For example, K\"{u}\c{c}\"{u}k et al.~\cite{kucuk2021improving} only focus on numerical assignments and predicates, while Liblit et al.~\cite{Liblit-isolation} locates the variables in return statements or on the left side of an assignment. To provide a fair comparison and make our results reproducible in future studies, we first provide a definition of \textit{fault-relevant variables} from the perspective of program repair.
Specifically, we define a variable (which can be a temporary variable of a predicate expression in the GSA form) as fault-relevant if it satisfies one of the following conditions:
\begin{enumerate}
    \item Variables that are directly modified (i.e., replaced or deleted) or inserted to the code for repairing the bug, such as the variable \codeIn{v} in the code change of ``\codeIn{v>0}$\rightarrow$\codeIn{v'>0}'' or the temporary predicate variable \codeIn{v=exp'} in the code change of ``\codeIn{if(exp||exp')\{\}}$\rightarrow$\codeIn{if(exp)\{\}}''.
    \item Variables whose values are directly affected by the newly inserted statement, such as the variable \codeIn{v} in the code change of ``\codeIn{v=exp;}$\rightarrow$\codeIn{if(cnd)\{v=exp;\}}''.
    \item If a data-flow-breaking statement (e.g., \codeIn{return}) is deleted or inserted, the temporary variable corresponding to the surrounding branch condition (if exists) since it is the indicator of the bug, such as variable \codeIn{v} in the code change of ``\codeIn{if(v=cnd)\{\}}$\rightarrow$\codeIn{if(v=cnd)\{return;\}}''
    \item If all the statements in the body of an \codeIn{if} statement are modified/deleted or an \codeIn{If} statement is deleted, indicating a special condition is incurred, the temporary variable of the condition, such as variable \codeIn{v} in the code changes of ``\codeIn{if(v=cnd)\{exp;\}}$\rightarrow$\codeIn{if(v=cnd)\{exp';\}}'' and ``\codeIn{if(v=cnd)\{exp;\}}$\rightarrow$\codeIn{exp;}''. Please note that if only a portion of statements are modified in the body, the failures are more likely to be caused by the incorrect statements themselves but unlikely related to the condition. In such cases, the first rule can be applied.
\end{enumerate}

The basic intuition of our definition is to locate variables that will be directly modified or are indicators that produce the bug. 
For example, the variables \codeIn{expPos} and \codeIn{str.length()} are both fault-relevant in the running example. Particularly, a buggy program may have multiple fault-relevant variables. On the basis of this definition, we have manually identified the fault-relevant variables for each bug used in our experiment, which will play as the ground truth and may also provide a standard for promoting future research (published in our open-source repository). Specifically, there are in average 4 fault-relevant variables per each bug in our studied dataset.
As it will be presented in Section~\ref{sec:rq3} that correctly locating these fault-relevant variables indeed can boost existing automatic program repair techniques, further demonstrating the reliability of the ground truth.


\subsubsection{Metrics}
\label{sec:metric}
Following previous studies~\cite{Ochiai,MUSE,li2021fault,lou2021boosting,zou2018empirical,Michael:SBFL}, we employ three metrics in the fault localization experiment.
\textbf{Recall at Top-N:} computes the number of bugs that have at least one fault-relevant variable correctly located within the Top-N position in the ranked list (aka., precision). We set N$\in\{1,3,5,10\}$ like existing studies~\cite{jiang2019combining,zou2018empirical}.
\textbf{Mean First Rank (MFR):} denotes the average rank of the first located fault-relevant variables for multiple bugs.
\textbf{Mean Average Rank (MAR):} When a bug has multiple fault-relevant variables, the MAR denotes the average rank of all these variables, while for multiple bugs, this metric denotes the average MAR of them. Following existing studies~\cite{zou2018empirical,jiang2019combining}, we adopt the average rank for variables in a tie. Please note that if the candidate variable list of an approach includes no fault-relevant variable, the corresponding bug will be removed when calculating the MAR and MFR for the approach to mitigate the bias of different predicates. Besides, we do not use the metric of \textit{Exam Score} used in statement-level fault localization~\cite{jiang2019combining,Michael:SBFL,zou2018empirical}. The reason is that it requires the total number of candidate variables in different approaches to the same for a fair comparison, which cannot be satisfied in our experiment.


In the application of patch filtering, we adopt two metrics following previous studies~\cite{xiong-icse18,xin2017identifying}, i.e., \textbf{Precision}=$N_{fi}/(N_{fi}+N_{fc})$ and \textbf{Recall}=$N_{fi}/(N_{fi}+N_{ni})$, where $N_{fi}$ denotes the number of incorrect patches filtered, $N_{fc}$ denotes the number of correct patches filtered, and $N_{ni}$ denotes the number of incorrect patches not filtered.




\section{Result Analysis}
\label{sec:result}

\subsection{RQ1: Overall Effectiveness of \tool{}}

\begin{figure}[t]
	\centering
	\includegraphics[width=0.98\columnwidth]{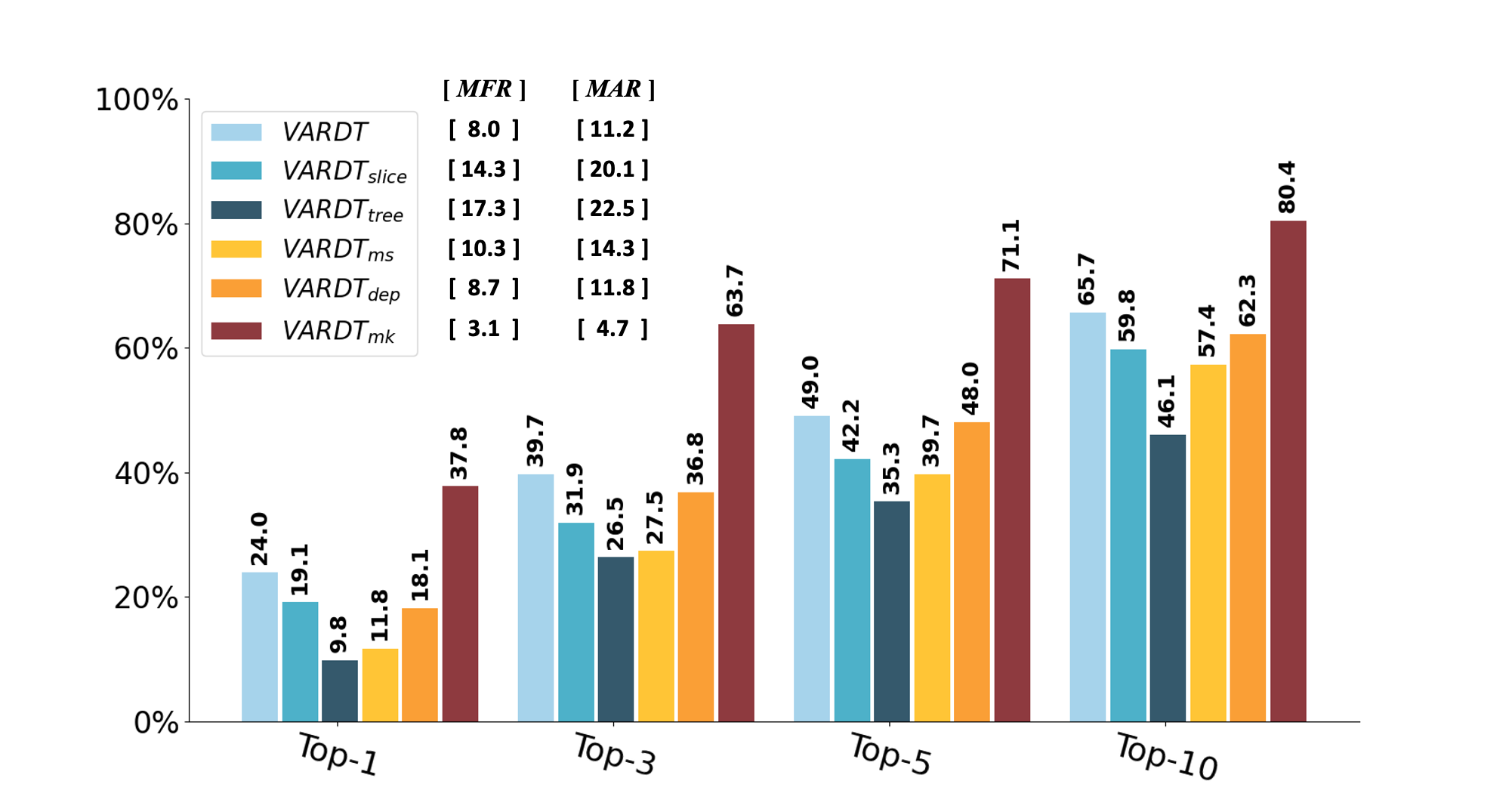}
	\caption{Results of \tool{} and its variants}
	\label{fig:ablation}
\end{figure}

As introduced, we conducted our experiment over 204 real-world bugs from Defects4J benchmark and compared the results with seven baseline approaches. Table~\ref{tab:othertools-topn} presents the experimental results of different approaches.
 From the table, we can see that our approach significantly outperforms the baselines. Specifically, \tool{} successfully located the desired fault-relevant variables for 24.0\%, 39.7\%, 49.0\% and 65.7\% of bugs within the Top-1/3/5/10 positions, respectively. The improvements over the baseline approaches range from 247.8\% to 515.4\% regarding the Top-1 recall, and the average improvements are 330.5\%.
Particularly, \tool{} significantly outperforms the latest state-of-the-art UniVal by 247.8\% with respect to the Top-1 recall. The results demonstrate that our approach is much more effective. 
Though effective, the absolute number of bugs located at Top-1 is still small (i.e., 24.0\%) for \tool{}. A major reason is the inaccuracy of the coarse-grained fault localization results used by \tool{}. As will be presented later (see Figure~\ref{fig:ablation}), when providing accurate method-level FL results, the effectiveness of \tool{} can be significantly improved.
Please note that the results of the compared approaches in our experiment are much worse than those results reported in the previous study~\cite{kucuk2021improving}. To ensure the correctness of the results, we further carefully checked them manually.
In addition, since the results were produced using the virtual machine published by the authors, we believe they should be reliable. We guess the decline may be caused by the different definitions of ground-truth variables, but they were not published by the authors, and thus we cannot reproduce their results. 


Furthermore, we also reported the detailed Top-1 results of different approaches on each project in Table~\ref{tab:othertools-top1}.
According to the results, \tool{} performs consistently well over different projects, and always outperforms the baselines, indicating the generalizability of our approach. 
Regarding the metrics of MAR and MFR shown in Table~\ref{tab:othertools-topn}, our approach also outperforms the competitors with at least 77.3\% and 71.5\% improvements, and the average improvements are 78.8\% and 73.6\%, respectively. The results further 
demonstrate the superiority of our approach. Please note that though \tool{} depends on the decision tree building process compared with baselines, \textbf{the cost of it is relatively small, i.e., 1.9s in average}.

\begin{table}[t]
\centering
\caption{Experimental result summary of all approaches} 
\label{tab:othertools-topn}
\resizebox{0.99\columnwidth}{!}{%
\setlength{\tabcolsep}{3pt}
\begin{tabular}{c|rrrrrrrrr}
\toprule
 \textbf{Metric}                                                     & \textbf{\tool{}}            & \textbf{UniVal}         & \begin{tabular}[c]{@{}l@{}}\textbf{Baah}\\ \textbf{-2010}\end{tabular} & \textbf{ESP}             & \begin{tabular}[c]{@{}l@{}}\textbf{NUMFL}\\ \textbf{-DLRM}\end{tabular} & \begin{tabular}[c]{@{}l@{}}\textbf{NUMFL}\\ \textbf{-QRM}\end{tabular} & \textbf{Ochiai}          & \begin{tabular}[c]{@{}l@{}}\textbf{Dstar}\\ \textbf{(Star=2)}\end{tabular} 
 \\ \midrule
\textbf{Top-1}  & \textbf{24.0\%} & 6.9\%  & 5.9\%   & 6.4\%  & 4.4\%     & 3.9\%    & 6.4\%  & 6.9\%        \\
\textbf{Top-3}  & \textbf{39.7\%} & 12.3\% & 10.3\%  & 9.8\%  & 8.8\%     & 11.8\%   & 14.2\% & 14.2\%       \\
\textbf{Top-5}  & \textbf{49.0\%} & 15.2\% & 15.2\%  & 12.3\% & 12.8\%    & 16.2\%   & 18.6\% & 18.1\%       \\
\textbf{Top-10} & \textbf{65.7\%} & 19.6\% & 18.1\%  & 15.7\% & 17.2\%    & 20.6\%   & 21.1\% & 20.6\% \\
\midrule
\textbf{MFR}  &  \textbf{8.0}  &  28.2  &  29.9  &  44.0  &  29.5  &  28.6  &  28.5   &  28.1  \\
\textbf{MAR}  &  \textbf{11.2}  &  49.3  &  53.2  &  65.6  & 52.0  &  49.5  & 52.5  &  51.4
\\ \bottomrule
\end{tabular}%
}
\end{table}
\begin{table}[t]
\centering
\caption{Top-1 results of all approaches}
\label{tab:othertools-top1}
\resizebox{\linewidth}{!}{%
\setlength{\tabcolsep}{3pt}
\begin{tabular}{l|rrrrrrrr}
\toprule
\textbf{Project}                                                     & \textbf{\tool{}}            & \textbf{UniVal}         & \begin{tabular}[c]{@{}l@{}}\textbf{Baah}\\ \textbf{-2010}\end{tabular} & \textbf{ESP}             & \begin{tabular}[c]{@{}l@{}}\textbf{NUMFL}\\ \textbf{-DLRM}\end{tabular} & \begin{tabular}[c]{@{}l@{}}\textbf{NUMFL}\\ \textbf{-QRM}\end{tabular} & \textbf{Ochiai}          & \begin{tabular}[c]{@{}l@{}}\textbf{Dstar}\\ \textbf{(Star=2)}\end{tabular} \\ \midrule
Compress        & \textbf{11.1\%} & \textbf{11.1\%} & \textbf{11.1\%} & 0.0\%           & \textbf{11.1\%} & 0.0\%           & \textbf{11.1\%} & \textbf{11.1\%} \\
Gson            & \textbf{42.9\%} & 14.3\%          & 14.3\%          & 0.0\%           & 0.0\%           & 0.0\%           & 14.3\%          & 14.3\%          \\
Codec           & \textbf{20.0\%} & 0.0\%           & 0.0\%           & 0.0\%           & 10.0\%          & 10.0\%          & 0.0\%           & 0.0\%           \\
Csv             & \textbf{20.0\%} & \textbf{20.0\%} & \textbf{20.0\%} & \textbf{20.0\%} & 0.0\%           & 0.0\%           & \textbf{20.0\%} & \textbf{20.0\%} \\
Lang            & \textbf{26.1\%} & 13.0\%          & 4.4\%           & 17.4\%          & 8.7\%           & 4.4\%           & 0.0\%           & 0.0\%           \\
JXml      & \textbf{25.0\%} & 0.0\%           & 0.0\%           & 0.0\%           & 0.0\%           & 0.0\%           & 0.0\%           & 0.0\%           \\
Chart           & \textbf{25.0\%} & 8.3\%           & 8.3\%           & 8.3\%           & 0.0\%           & 0.0\%           & 0.0\%           & 0.0\%           \\
JCore     & \textbf{6.3\%}  & 6.3\%           & 0.0\%           & 0.0\%           & 0.0\%           & 0.0\%           & 6.3\%           & 6.3\%           \\
Jsoup           & \textbf{16.7\%} & 5.6\%           & 0.0\%           & 5.6\%           & 5.6\%           & 0.0\%           & 0.0\%           & 0.0\%           \\
JxPath          & \textbf{12.5\%} & 0.0\%           & 0.0\%           & 0.0\%           & 0.0\%           & \textbf{12.5\%} & 0.0\%           & 0.0\%           \\
Math            & \textbf{26.1\%} & 13.0\%          & 13.0\%          & 17.4\%          & 13.0\%          & 13.0\%          & 13.0\%          & 13.0\%          \\
JDatabind & \textbf{32.1\%} & 3.6\%           & 10.7\%          & 3.6\%           & 0.0\%           & 0.0\%           & 17.9\%          & 17.9\%          \\
Time            & \textbf{25.0\%} & 0.0\%           & 8.3\%           & 0.0\%           & 0.0\%           & 8.3\%           & 0.0\%           & 8.3\%           \\
Cli             & \textbf{37.5\%} & 12.5\%          & 0.0\%           & 12.5\%          & 12.5\%          & 12.5\%          & 0.0\%           & 0.0\%           \\
Mockito         & \textbf{28.6\%} & 0.0\%           & 0.0\%           & 0.0\%           & 0.0\%           & 0.0\%           & 4.8\%           & 4.8\%           \\
\midrule
\textbf{TOTAL}  & \textbf{24.0\%} & \textbf{6.9\%}  & \textbf{5.9\%}  & \textbf{6.4\%}  & \textbf{4.4\%}  & \textbf{3.9\%}  & \textbf{6.4\%}  & \textbf{6.9\%} 
\\ \bottomrule
\end{tabular}%
}
\end{table}
\begin{figure}[tb]
	\centering
	\includegraphics[width=0.85\columnwidth]{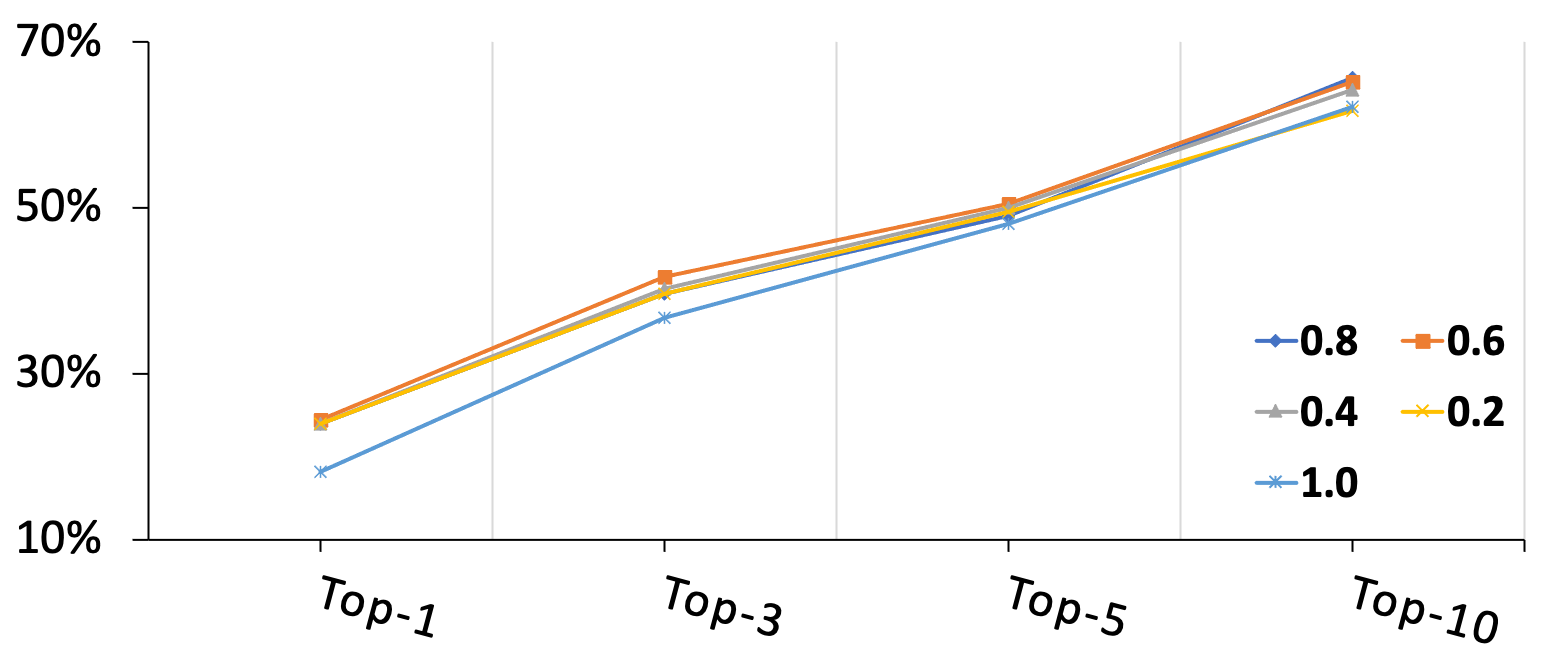}
	\caption{Effects of different values of dependency factor}
	\label{fig:depfactor}
\end{figure}

\subsection{RQ2: Contribution of Each Component}

In order to evaluate the effectiveness of each component in \tool{}, we have conducted an ablation study with a set of variants of \tool{}, which have been introduced in Section~\ref{sec:baseline}. 
Figure~\ref{fig:ablation} presents the results of each variant regarding the metrics of Top-N recall, MFR and MAR.
According to the results, all components in \tool{} largely contributed to the overall effectiveness of \tool{} since a large drop on the Top-N recall was incurred when removing any one of them. Specifically, regarding the metric of Top-1 recall, the dynamic program slicing contributed 25.7\% higher effectiveness (\textit{vs} \variant{slice}), the tree model contributed 144.9\% (\textit{vs} \variant{tree}), the \textit{dependency penalty} contributed 32.6\% (\textit{vs} \variant{dep}), and the use of method score for variable ranking contributed 103.4\% (\textit{vs} \variant{ms}), respectively. However, all of them always outperform the baseline approaches.
In particular, the core novel component (i.e., tree model) in \tool{} makes the largest contribution. In summary, the ranking of component contributions is \textit{tree model $>$ method score $>$ dependency penalty $>$ program slicing} regarding Top-N.

Since the method score largely affects the effectiveness of \tool{}, a question may naturally raise: Whether \tool{} can be further improved by providing a more accurate fault localization result (e.g., providing the genuine faulty method). Therefore, we empirically evaluated the fault localization results of \tool{} in the circumstance where the faulty method was known, for which we created another variant \variant{mk}. The results of \variant{mk} are also presented in Figure~\ref{fig:ablation}.
 \variant{mk} successfully located the desired variables for 37.8\% of bugs at Top-1, the improvements over \tool{} are about 57.5\%. Moreover, the Top-10 recall is as high as 80.4\%, indicating the promise of incorporating a more effective method-level fault localization technique into \tool{}.

Then, we further investigated the impact of the configuration for \depfactor{}, which represents the strength of \textit{dependency penalty}. According to Formula~\ref{eq:dep}, the smaller the value is, the larger the penalty will be (i.e., the variable will be less likely to be selected). 
Figure~\ref{fig:depfactor} presents the results when taking different values, where 0.8 is the default value. From the figure, we can see that \tool{} achieved the best overall result when taking the value in $[0.6, 0.8]$, and the impact of this configuration is relatively small. Specifically, when taking 0.6, \tool{} achieved the highest Top-1 recall as 24.5\%, whereas it achieved the lowest Top-1 recall as 18.1\% when taking 1.0. The result indicates that \tool{} is insensitive to this configuration although it is indeed important according to the result of \variant{dep}, which completely removes the \textit{dependency penalty} component. 

Finally, we also investigated the impact of program slicing in depth on the \textbf{space reduction} of candidate variables. The result shows that the reduction ratio ranges from 18.5\% to 42.2\% over different projects, and in average is 31.9\%, denoting the necessity of it for improving efficiency.

\newcommand{\var}[1]{\textsc{DT}$_\mathrm{\textit{#1}}$}
\begin{table}[t]
	\centering
	\caption{Experimental results in patch filtering}
	\label{tab:patchfilter}
	\resizebox{\columnwidth}{!}{%
	\begin{tabular}{lc|rrrrr}
		\toprule
		\textbf{Project} & \textbf{\#All(Correct)} & \textbf{\var{Top-1}}           & \textbf{\var{Top-3}}           & \textbf{\var{Top-5}}         & \textbf{\var{Top-10}}          & \textbf{PATCH-SIM}   \\ \midrule
		Math     & 24(3)                                & 22(2)            & 22(2)            & 20(1)            & 14(0)            & 15(0)   \\
		Lang    & 10(2)                                 & 10(2)            & 10(2)            & 10(2)            & 9(1)             & 2(0)   \\
		Chart   & 14(2)                               & 14(2)            & 8(1)             & 7(1)             & 6(1)             & 4(0) \\
		Time  &  8(1)                               & 7(1)             & 6(0)             & 6(0)             & 6(0)             & 6(0)   \\
		Mockito  & 2(1)                                & 1(0)             & 1(0)             & 1(0)             & 1(0)             & -    \\ \midrule
		\textbf{TOTAL}  &58(9)& \textbf{54(7)}   & \textbf{47(5)}   & \textbf{44(4)}   & \textbf{36(2)}   & \textbf{27(0)}  \\ \midrule
		\multicolumn{2}{c|}{\textbf{Precision}} & \textbf{87.0\%} & \textbf{89.4\%} & \textbf{90.9\%} & \textbf{94.4\%} & \textbf{100.0\%}   \\
		\multicolumn{2}{c|}{\textbf{Recall}} & \textbf{95.9\%} & \textbf{85.7\%} &  \textbf{81.6\%} &  \textbf{69.4\%} & \textbf{55.1\%}    \\
		\bottomrule
	\end{tabular}%
}
\end{table}

\subsection{RQ3: Performance in Patch Filtering}
\label{sec:rq3}

To evaluate whether our finer-grained variable-level fault localization results can further the effectiveness of downstream APR techniques, we adapted \tool{} to the task of patch filtering and compared the result with the state-of-the-art PATCH-SIM. The details have been introduced in Section~\ref{sec:baseline}. 

Table~\ref{tab:patchfilter} presents the experimental results. Specifically, we list the number of all plausible and correct patches per each project in the left part of the table, while list the filtering results in the right part. Particularly, \var{Top-N} represents that we use the Top-N variables located by \tool{} to filter patches. 
In each cell, \textit{X(Y)} denotes the corresponding approach in total filtered \textit{X} patches, in which \textit{Y} patches were correct patches. According to the result, although our approach was not designed as a comprehensive and standalone patch filtering technique, it still could filter out about 69.4\% incorrect patches using the Top-10 results of \tool{}, while PATCH-SIM only filtered 55.1\%. That is, \tool{} outperforms PATCH-SIM by 26.0\% in terms of incorrect patches filtered. Particularly, the patch precision (the percentage of correct patches over all patches) increased to 31.8\% and 29.0\% from 15.5\% by \var{Top-10} and PATCH-SIM respectively after filtering.
The result indicates the performance of \tool{} and the feasibility of boosting automatic program repair techniques by filtering incorrect patches using a finer-grained fault localization. It also reflects the reliability of our ground truth since it is indeed closely related to the repair of the bug. Besides, designing new automatic program repair techniques based on the variable-level fault localization potentially can further improve the number of correct patches since many incorrect patches can be avoided to be generated in the online repair process, and thus correct patches will have more possibility to be generated. More studies can be conducted in this direction.

Though effective, our approach tends to incorrectly filter out correct patches compared with PATCH-SIM. For example, two correct patches were filtered out by \var{Top-10} while none by PATCH-SIM. Particularly, with the decrease of variable numbers (i.e., from Top-10 to Top-1), although more incorrect patches can be filtered out, the \textit{precision} of filtering will also sharply drop. When using the Top-1 result (i.e., only one candidate variable for each bug), 7 out of 9 correct patches will be filtered due to the inaccuracy of \tool{}. Particularly, after further analyzing the results of the two approaches, we found that there were only 19 incorrect patches that were commonly filtered out by both \var{Top-10} and PATCH-SIM. In other words, (34+27)-19=42 incorrect patches and 2 correct patches could be filtered by combining these two, leaving the \textit{precision} and \textit{recall} of filtering as 95.5\% and 85.7\%, respectively, and the patch precision will also increase to 50\%. The result reflects that they complement each other.




\section{Discussion}
\label{sec:diss}
\noindent \textbf{Limitation}: As explained in Section~\ref{sec:subject}, \tool{} requires that at least three test cases (including at least one failed and one passed) cover the faulty method,
which may affect the usability of our approach in practice since the accompanied test suite tend to be weak~\cite{qi15,xiong-icse18}. 
In such cases, the state-of-the-art test generation approaches~\cite{Fraser2011EvoSuiteAT,zhang2021partial,AFL++} may be potentially combined to overcome this limitation.


\noindent \textbf{Internal threats}: The threats to internal validity mainly lie in the implementation and ground truth used in our experiment. In order to ensure the reliability of our implementation, two authors have carefully checked its correctness through code review, which can mitigate this threat. Regarding the ground truth, we have provided a clear definition of fault-relevant variables, based on which we manually analyzed the source code. Therefore, we believe it is reliable. Additionally, the evaluation result of \tool{} in the patch filtering application also improves our confidence. Finally, we have published all our data and implementation to ease the replication.


\noindent \textbf{External threats}: The threats to external validity mainly lie in the used subjects. In our experiment, we only adopted a subset of the bugs from the Defects4J benchmark according to the constraints introduced in Section~\ref{sec:subject}. However, since the studied bugs are from 15 different real-world projects, we believe it can be representative to some extent. The effectiveness of \tool{} on a wider range of projects beyond Defects4J remains to be studied.

\noindent \textbf{Future Work}: As can be observed, the fault-relevant variables and their constraints indeed can provide possible hints for program repair according to the example shown in Figure~\ref{fig:tree-pic} and the results in patch filtering. In the future, we plan to further investigate the performance of \tool{} for assisting human developers in the manual repair process.
\section{Related Work}
\label{sec:related}

\subsection{Variable-based Fault Localization}


Our approach targets the variable-based fault localization, the most related techniques are UniVal~\cite{kucuk2021improving}, NUMFL~\cite{bai2015numfl}, ESP~\cite{core2021statistical}, and Baah2010~\cite{Baah2010causal}, which have been introduced in Section~\ref{sec:baseline}. Different from them, our approach locates fault-relevant variables by leveraging decision trees to build variable constraints for discriminating failed and passed test runs, which is the first time as far as we are aware.
Besides, the statistical debugging~\cite{Liblit-sampling} and its following work~\cite{Zheng:SD:multi-bug,compound-predicate,Jiang-sd-cfp,HOLMES,sober-TSE,core2021statistical} are also related to our approach, which depends on test coverage to compute the importance of a set of pre-defined predicates. 
On the contrary, our approach uses a dependency-enhanced tree model to identify fault-relevant variables, but not simply their coverage. In addition, existing studies also employed decision tree~\cite{sbfl-dt} or random forest models~\cite{xai4fl} in fault localization. However, they were designed for improving the statement-level fault localization, whereas our approach targets the variable level and additionally incorporates the dependency factor for model building.

Besides locating fault-relevant variables directly, several studies use variable/value profiles to boost statement-level fault localization. For example, a set of studies devoted to improving statement-level fault localization by replacing the values of certain expressions with alternative values in order to make the failed test pass~\cite{zhang2006locating,jeffrey2008fault,chandra2011angelic}.
Recent studies~\cite{MUSE,Metallaxis} incorporated mutation analysis to improve fault localization. Shen et al.~\cite{vunloc} combined statistical localization with directed fuzzing to overcome the over-fitting and estimation bias problem in fault localization.
Different from them, our work aims at locating the finer-grained fault-relevant variables directly. 

Finally, there are also some interventional fault localization approaches depending on variable values~~\cite{zeller-simplifying,zeller-cause-effect}, Compared with them, our approach is fully automatic. The latest studies also employed different models to combine the strength of multiple techniques~\cite{jiang2019combining,Xuan:2014:LCM:2705615.2706097,zou2018empirical,li2021fault}. These techniques can be further combined with our approach and boost its effectiveness by providing a more precise method-level fault localization result. In turn, our approach may also improve existing techniques by integrating it into them.


\subsection{Automatic Patch Filtering}

In order to improve the patch quality (i.e., precision) in automatic program repair, researchers have proposed a series of patch filtering techniques. Among them, test-generation-based techniques are the most widely studied, and the core challenge is the lack of test oracles. Facing this challenge, existing studies employed different strategies. Yang et al.~\cite{Yang17} proposed Opad, which filters patches that cause program crashes or produce memory errors. Xin and Reiss~\cite{xin2017identifying} proposed DiffTGen which depends on human experts to provide the test oracle. While Xiong et al.~\cite{xiong-icse18} proposed PATCH-SIM that estimates the test results by measuring the execution similarity of test cases before and after repair. On the contrary, Tan et al.~\cite{tananti} pre-defined a set of anti-patterns that easily produce incorrect patches for patch filtering. Recently, Ye et al.~\cite{ye2021ods} proposed to employ a machine learning method to classify the correctness of patches, while Wang et al.~\cite{wang2020automated} proposed a deep-learning-based approach. Different from existing approaches, our work focuses on improving the fault localization effectiveness by providing finer-grained results, which can also aid the patch filtering process but from a different perspective, and thus is orthogonal to them.

\section{Conclusion}
\label{sec:conclude}

In this paper, we have proposed a variable-level fault localization technique, named \tool{}, in which we designed a novel program-dependency-enhanced decision tree model to aid the identification of fault-relevant variables. We have evaluated the effectiveness of \tool{} in both fault localization and patch filtering applications by comparing with the state-of-the-art techniques. The results demonstrate that \tool{} significantly outperformed the baseline approaches, 
where the improvements are at least 247.8\% and 26.0\% regarding bugs located within Top-1 and the number of incorrect patches filtered, respectively in the aforementioned applications, demonstrating the effectiveness of our approach.


\balance
\bibliographystyle{IEEEtran}
\bibliography{ref}{}

\end{document}